\documentclass[journal=jacsat,manuscript=article]{achemso}

\usepackage[version=3]{mhchem} % Formula subscripts using \ce{}
\usepackage{amssymb}
\usepackage{multirow}
\usepackage[T1]{fontenc}
\usepackage{hyperref}
\usepackage{xcolor}
\newenvironment{revised}{%
  \color{black}%
}{}
\author{Joshua Zhi En Tan}
\affiliation[NTU]
{Division of Mathematical Sciences, School of Physical and Mathematical Sciences, Nanyang Technological University, Singapore 637371}

\author{JunJie Wee}
\affiliation[UMich]
{Michigan State University, Department of Mathematics, East Lansing, MI 48824, USA}
\email{weejunji@msu.edu}

\author{Xue Gong}
\email{xue.gong@ntu.edu.sg}

\author{Kelin Xia}
\email{xiakelin@ntu.edu.sg}
\affiliation[NTU]
{Division of Mathematical Sciences, School of Physical and Mathematical Sciences, Nanyang Technological University, Singapore 637371}
\title[Topology-enhanced machine learning model (Top-ML) for anticancer peptides prediction]{Topology-enhanced machine learning model (Top-ML) for anticancer peptides prediction}

\abbreviations{IR,NMR,UV}

\begin{document}


\begin{abstract}
  Recently, therapeutic peptides have demonstrated great promise for cancer treatment. To explore powerful anticancer peptides, artificial intelligence (AI)-based approaches have been developed to systematically screen potential candidates. However, the lack of efficient featurization of peptides has become a bottleneck for these machine-learning models. In this paper, we propose a topology-enhanced machine learning model (Top-ML) for anticancer peptides prediction. Our Top-ML employs peptide topological features derived from its sequence ``connection'' information characterized by vector and spectral descriptors. Our Top-ML model, employing an Extra-Trees classifier, has been validated on the AntiCP 2.0 and mACPpred 2.0 benchmark datasets, achieving state-of-the-art performance or results comparable to existing deep learning models, while providing greater interpretability. Our results highlight the potential of leveraging novel topology-based featurization to accelerate the identification of anticancer peptides.
\end{abstract}

\section*{Introduction}
Despite the extensive research into the prevention and diagnosis of cancer over the past few decades, cancer remains one of the leading causes of death worldwide. Statistics from the World Health Organization \cite{WHO2021} show that cancer accounted for an estimated 10 million deaths in 2020, representing nearly one in six deaths. Traditional approaches to cancer treatment often come with significant limitations, such as systemic toxicity and drug resistance. Traditional cancer treatment relies on radiotherapy and chemotherapy; however, both methods have downsides, as they target both normal and cancerous cells. Furthermore, highly adaptable cancer cells can develop drug resistance, further reducing the efficacy of these treatments.

In recent years, there has been a growing interest in anticancer peptides (ACPs) as a novel alternative treatment for cancer. ACPs are a specific type of antimicrobial peptides (AMP) typically consisting of less than 50 amino acids. They differ from other AMPs by their cationic nature and low molecular weight \cite{xie2020anticancer}. Peptide therapy offers significant advantages over traditional cancer treatments due to its high specificity, low toxicity, superior membrane penetration ability, and easier chemical modifications. However, the experiment-based discovery and design of ACPs are extremely costly, time-consuming, and labor-intensive, which leads to poor scalability. To address these challenges, researchers are increasingly turning to machine learning for the screening and design of ACPs.

Featurization or feature engineering is key to the performance of machine learning models for ACPs design and discovery. The ACPs identification task was formulated as a binary classification machine learning problem in these studies \cite{feng2022me, lv2021anticancer, agrawal2021anticp, charoenkwan2021improved, liu2022antimf, ahmed2021acp}. In these methods, features are constructed from the peptides and then used to train ML models. Traditional molecular descriptors are primarily designed based on sequence composition, such as amino acid composition, dipeptide composition, and terminus composition to establish strong structure-function relationships \cite{agrawal2021anticp}. In addition, other classes of physicochemical properties, including hydrophobicity and amphipathicity, have been utilized as feature encoding schemes \cite{liang2021large}. The effectiveness of machine learning techniques heavily depends on molecular featurization and representations \cite{nguyen2020review, cang:2018representability, chuang2020learning}.

Recently, topological data analysis (TDA) \cite{Edelsbrunner:2002, Zomorodian:2005}, which is a model deeply rooted in algebraic topology, has demonstrated its great power in molecular representation and featurization \cite{cang:2017topologynet}. In particular, TDA-based machine learning models have achieved great successes in various steps of drug design, including protein-ligand binding affinity prediction \cite{cang:2017topologynet}, protein stability change upon mutation prediction \cite{cang:2017analysis}, toxicity prediction \cite{wu:2018quantitative}, solvation free energy prediction \cite{wang2016automatic}, partition coefficient and aqueous solubility \cite{wu2018topp}, binding pocket detection \cite{zhao2018protein}, protein mutation analysis \cite{wang2020topology}, and drug discovery \cite{gao2020generative}. In the D3R Grand Challenge, a worldwide competition for computational models in drug design, TDA-based models have consistently delivered state-of-the-art results \cite{nguyen2018mathematical}. Further, TDA-based models have also achieved great successes in material data analysis \cite{nakamura2015persistent, hiraoka2016hierarchical}. The key idea of these models is to characterize the structural ``connection'' information with topological tools.

Here, we propose a topology-enhanced machine learning model (Top-ML) for ACPs prediction. The key idea is to use topological features to characterize the ``connection'' pattern between amino acids within peptide sequences. More specifically, we employ vector features and spectral features derived from amino acid sequences to describe these connection properties, i.e., how amino acids, amino acid pairs, and amino acid groups relate to each other. Our Top-ML considers four types of peptide features, namely natural vector \cite{tseng1998natural}, Magnus vector \cite{wu2019magnus},  terminal composition feature, and spectral feature. By combining all four feature types with an Extra-Trees classifier, we demonstrate that the proposed Top-ML achieves state-of-the-art results on the alternative dataset of AntiCP 2.0 and accuracy comparable to the best deep learning models on the main dataset of AntiCP 2.0 and mACPpred 2.0. 
Additionally, our model offers greater interpretability, which we demonstrate through the feature importance analysis shown in Figure \ref{fig:importance} using the mACPpred 2.0 dataset.

\section*{Results}

\subsection*{Peptide vector representations}

\paragraph{Natural Vector}
Natural vector has been widely used in the construction of phylogenetic trees \cite{wen2014kmer, zhang2019phylogenetic, zhao2022protein}. Here we use natural vector representation for peptide sequences. The natural vector was first conceived for the analysis of genomes and has been previously applied to DNA sequences \cite{li2017novel}. Consider a peptide sequence $\mathcal{S} = x_1x_2\cdots x_n$ of length $n$, where $x_i$ lies in the set of 20 unique amino acids $\{A, C, D, \cdots, Y\}$. For each peptide sequence $\mathcal{S}$ and amino acid $R$, we compute three statistical metrics: $n_R$, representing the number of times $R$ appears in $\mathcal{S}$; $\mu_R$, the average position of $R$; and $D_2^R$, the variation of position. The natural vector concatenates the three statistics calculated for each unique amino acid, forming a vector of length 60, i.e., ($n_A$, $\mu_A$, $D_2^A$, ... $n_R$, $\mu_R$, $D_2^R$, ... $n_Y$, $\mu_Y$, and $D_2^Y$). The natural vector encodes the presence and location of amino acids which correlate with anticancer properties. For instance, a previous study found that A, F, H, K, L, and W are more prevalent in ACPs, whereas C, G, R, and S are more common in non-ACPs \cite{agrawal2021anticp}.

\paragraph{Magnus Vector} In combinatorial group theory, Wilhelm Magnus studied representations of free groups by non-commutative power series. For a free group $F$ with basis $x_1, \cdots, x_n$ and a power series ring $\prod$ in indeterminates $\xi_1, \cdots, \xi_n$, Magnus proved that the map $\mu: \xi_i \mapsto 1 + \xi_i$ defines an isomorphism from $F$ into the multiplicative group $\prod^\times$ of units in $\prod$ \cite{lyndon1977combinatorial}. Magnus's work laid the foundation for the development of Magnus representation, which has been recently applied to the analysis of DNA and RNA sequences in genome analysis \cite{wu2019magnus}.

Peptide sequence-based Magnus representation and Magnus vector can be developed and used for the characterization of peptide properties. The Magnus representation of a peptide sequence $\mathcal{S}$, denoted by $\rho(\mathcal{S})$, is defined as the product $\rho(\mathcal{S}) = \prod_{i=1}^n (1+x_i)$ in the non-commutative polynomial algebra $R\langle A, C, D, \cdots, Y\rangle$, where $R$ is a commutative ring. In this paper, we use $R = \mathbb{Z}$. For example, the peptide sequence $\mathcal{S} = ACD$ will be mapped to $\rho(\mathcal{S}) = (1+A)(1+C)(1+D) = 1+A+C+D+AC+AD+CD+ACD$.

In implementation, the Magnus vector of a peptide sequence $S$, denoted as $v(S)$, is obtained by the following steps. First, we arrange the set of possible words over the amino acids $\{A, C, D, \cdots, W, Y\}$ of length less than or equal to $n$, first by ascending order of length and then by lexicographic order. With respect to the above arrangement, assign each term present in $\rho(S)$ with a constant coefficient $c \in R$, and 0 for each term not present in $\rho(S)$. In our model, we use $c = 1$ and only consider subsequences with a maximum length of 2. As each peptide sequence can contain a maximum of 20 unique amino acids \cite{agrawal2021anticp}, there are a maximum of $\sum_{i=1}^2 20^i = 420$ possible subsequences of up to length 2, i.e., $A, C, D, \cdots, W, Y, AA, AC, AD, \cdots, YW, YY$. 

Consider the peptide sequence $\mathcal{S} = ACD$. The subsequences present in $\mathcal{S}$ are $A, C, D, AC,$ and $CD$. The corresponding Magnus vector $v(S) = (1, 1, 1, 0, ..., 0)^T$ is a 420-dimensional binary vector with 1s at the positions corresponding to these 5 subsequences and 0s elsewhere. Note that due to the non-commutativity of the variables, if $S' = ADC$, we observe that $\rho(S') \neq \rho(S)$ and $v(S') \neq v(S)$. From our numerical experiments detailed in the Results section, we found that the Magnus vector of length 420 are sufficient to achieve state-of-the-art results in ACPs classification.

In our implementation, we consider the Magnus vectors for non-overlapping $k$-mers for each peptide sequence. A $k$-mer represents a segment of $k$ consecutive amino acids within a given peptide sequence. We compute the Magnus vector for each $k$-mer, selected using non-overlapping sliding windows of size $k$ to avoid redundancy and reduce computational cost. The sliding window begins at the first index, and we truncate the tail if the remaining sequence is shorter than $k$. Therefore, the number of $k$-mers for a peptide sequence of length $n$ is the largest integer less than or equal to $\frac{n}{k}$.  We then form a mean Magnus vector $m(S)$ by averaging the Magnus vectors across all $k$-mers, capturing both $k$-mer frequency information and internal subsequence structure. We experiment with sliding window sizes $k=4, 5, 6, 7$ and find that $k=5$ yields good prediction performance for AntiCP 2.0 Datasets, as detailed in Table \ref{tbl:MagnusA} in the Methods section. This result aligns with the observation of minimal classification errors for the Baltimore and genus classification labels reported in \cite{huang2016ensemble}. Therefore, we selected $k = 5$ for our final Top-ML model.

\paragraph{Terminal Composition Features} 
Previous research into peptide representation has found that terminal residues provide vital information about the biological and physiological functions of peptides \cite{otvos2005antibacterial}. The terminus composition provides local sequence information in contrast to the global sequential information when the full length of the peptide is used for featurization. We compute the Magnus and natural vectors based on the terminus compositions of peptides (i.e., N5, C5, N5C5, $\cdots$, C15, N15C15). For example, extracting the first 5 residues from the N-terminus is termed N5. Our experiments show that the C15 and N15C15 terminus provided the best prediction accuracy for AntiCP 2.0 Datasets A and B respectively, as shown in Table \ref{tbl:terminalA}.

\subsection*{Peptide spectral representation}
In this section, we introduce a peptide spectral representation, which is inspired by the spectral theory of combinatorial Laplacian matrices. Simplicial complexes offer an effective representation for capturing the underlying structural information and its interactions. We refer the reader to the Method section for more information regarding simplicial complexes and their combinatorial Laplacian matrices. Using simplicial complexes as a molecular structural representation has generated tremendous success for combinatorial Laplacian matrices in applications such as drug design and discovery \cite{meng2021persistent}. However, unlike molecular structures, which inherently possess 3D structures, peptide data is sequential and cannot be directly represented by various topological objects such as graphs and simplicial complexes. Therefore, we need to first construct simplicial complexes from peptide sequences.
 
In order to construct a peptide spectral representation, we first define a sequence-based boundary matrix. For any peptide sequence $S$ of length $n$, we denote the set of possible subsequences of length $k$ as $S_k$, where $1 \leq k \leq n$. Then the sequence-based boundary matrix $\hat{\mathbf{B}}_k$ ($1 \leq k \leq n$) is an $|S_{k}| \times |S_{k+1}|$ matrix constructed from the relationships between subsequences $w^{k} \in S_{k}$ and $w^{k+1} \in S_{k+1}$. For a peptide sequence $S$ of length $n$, its $k$-th sequence-based boundary matrix $\hat{\mathbf{B}}_k$ is defined as follows:
\begin{equation}
\hat{\mathbf{B}}_k(i,j) = \begin{cases}
f(w_i^{k}), & \text{if } w_i^{k} \preceq w_j^{k+1},\\
0 & \text{otherwise },
\end{cases}
\label{eqn:Bk}
\end{equation}
where $f(w)$ is a real-valued function, $w_i^{k} \preceq w_j^{k+1}$ means that the $w_i^{k}$ is a subsequence of $w_j^{k+1}$. 
The simplest choice is setting $f(w) = 1$, in which case we obtain the standard boundary matrix in Equation (\ref{eqn:Bk}), where entry $\hat{\mathbf{B}}_k(i,j)$ is 1 if $w_i^{k}$ appear as a subsequence of $w_j^{k+1}$ for at least once, and 0 otherwise. However, since a peptide sequence consists of 20 possible amino acid types, this would result in many different sequences having identical boundary matrices. To enhance expressive power and better differentiate peptide sequences, we incorporate both frequency and positional information by considering two definitions for $f(w)$. 

\begin{figure*}[]
	\centering
	\includegraphics[width=\textwidth]{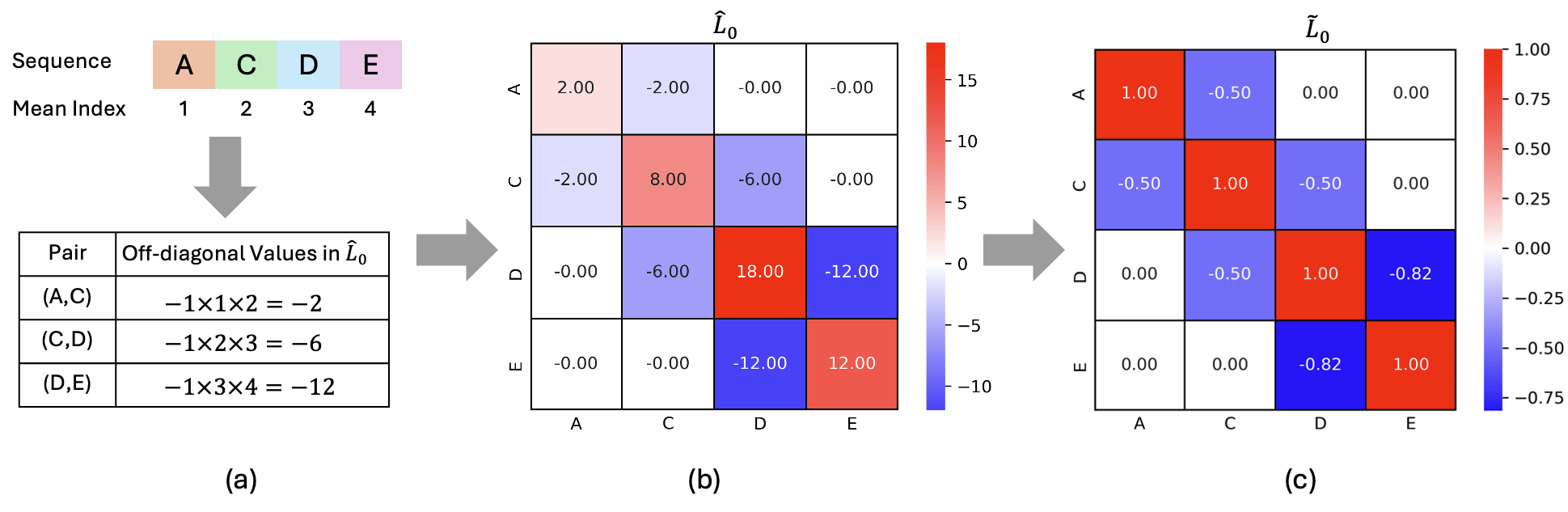}
	\caption{Illustration of the transformation of (a) the peptide sequence ``ACDE'' into (b) a sequence-based Laplacian $\hat{\mathbf{L}}_0$ and (c) its corresponding symmetric normalized Laplacian matrix $\tilde{\mathbf{L}}_0$ based on the mean index positions of the amino acids. The largest off-diagonal entry in (b) $\hat{\mathbf{L}}_0$ in terms of absolute value corresponds to the amino acid pair (D, E) since D and E have the highest mean index positions. After normalization, the diagonal entries in (c) $\tilde{\mathbf{L}}_0$ become ones, and the off-diagonal entries take values between -1 and 0.}
    \label{fig: Laplacian}
\end{figure*}

The first defines $f(w)$ as the number of occurrences of the subsequence $w$ within the sequence $S$. For instance, in the sequence ``FFSFS'', the subsequence ``FS'' occurs twice. Another definition of $f(w)$ is the mean index position of the first amino acid in $w$ across all its occurrences within sequence $S$. For example, in ``FFSFS'', the index positions of ``FS'' are 2 and 4, resulting in a mean index position of 3. Figure \ref{fig: Laplacian} shows another example for the construction of the sequence-based Laplacian using the mean index position. The choice of $f(w)$ for the final model is the mean index position based on the  model performance as presented in Table \ref{tbl:SpectPepA}.
\begin{revised}

\end{revised}

Similar to the combinatorial Laplacian defined on simplicial complexes, we compute our sequence-based Laplacian matrices $\hat{\mathbf{L}}_k$ using Equation (\ref{eq:sequnce_laplacian}) below:
\begin{equation}
\hat{\mathbf{L}}_k = \begin{cases}
\hat{\mathbf{B}}_1\hat{\mathbf{B}}_1^T, & \text{if } k=0 \\
\hat{\mathbf{B}}_k^T\hat{\mathbf{B}}_k + \hat{\mathbf{B}}_{k+1}\hat{\mathbf{B}}_{k+1}^T, & \text{if } k \geq 1.
\end{cases}
\label{eq:sequnce_laplacian}
\end{equation}
For $\hat{\mathbf{L}}_k$ to be positive semi-definite, we update the diagonals of $\hat{\mathbf{L}}_k$ using the sum of its off-diagonals and then assign all off-diagonal values a negative sign. This ensures the sum of every row in $\hat{\mathbf{L}}_k$ is zero. We then convert $\hat{\mathbf{L}}_k$ to its symmetric normalized version $\tilde{\mathbf{L}}_k$ as defined in Equation (\ref{eq:normalized L}).

Let's consider $\hat{\mathbf{B}}_1$ as an example. Given that all peptides consist of a maximum of 20 unique amino acids \cite{agrawal2021anticp}, the rows in $\hat{\mathbf{B}}_1$ represent these 20 amino acids (A, C, D, \ldots, Y), while the columns represent the 400 pairs of amino acids (AA, AC, AD, \ldots, YY). Therefore, $\tilde{\mathbf{L}}_1$ is a $20 \times 400$ matrix, $\tilde{\mathbf{L}}_0$ is a $20 \times 20$ matrix, and $\tilde{\mathbf{L}}_1$ is a $400 \times 400$ matrix. This resulting vector containing eigenvalues is used as input features for downstream machine learning models. $\tilde{\mathbf{L}}_0$ produces a feature vector of size 20 while $\tilde{\mathbf{L}}_1$ produces a feature vector of size 400. The two feature vectors are concatenated to form a feature vector of size 420.

\begin{revised}
Our Laplacian can be viewed as a weighted Laplacian matrix, where the weights reflect either the frequency or position at which subsequence pairs appear together. $\tilde{\mathbf{L}}_0$ resembles a graph Laplacian where nodes represent amino acids, and edges indicate consecutive appearances. Similarly, in $\tilde{\mathbf{L}}_1$, nodes represent dipeptides (or 2-mers), with edges connecting dipeptides that are adjacent and overlap by one amino acid. Such $k$-mers representation are widely used in bioinformatics to cluster amino acid sequences \cite{steinegger2018clustering}. The degeneracy of the zero eigenvalue counts the number of disconnected components of the underlying graph, while a smaller $p$-th eigenvalue indicates better $p$-way clustering \cite{lee2014multiway}. In our setting, good clustering means that subsequences within the same cluster frequently appear together, while those from different clusters are more separated. Biologically, such clustering may reflect amino acid groupings based on physicochemical properties, such as polar, neutral, or hydrophobic. Previously, adjacency pattern between these groups (referred to as Transition) have been used to predict protein folding \cite{dubchak1995prediction} and function \cite{cai2003svm}.
\end{revised}

\begin{revised}
    Compared to traditional sequence-based features, such as Amino Acid Composition (AAC) and Dipeptide Composition (DPC) \cite{agrawal2021anticp}, which mainly capture the frequency of individual amino acids or amino acid pairs, our spectral features incorporate additional structural information, including the adjacency between amino acid pairs, their positional distribution, and clustering behavior, offering a more expressive sequence characterization.
\end{revised}

\subsection*{Topology-enhanced Machine Learning (Top-ML)}

\begin{figure}[t!]
	\centering 
	\includegraphics[width=\linewidth]{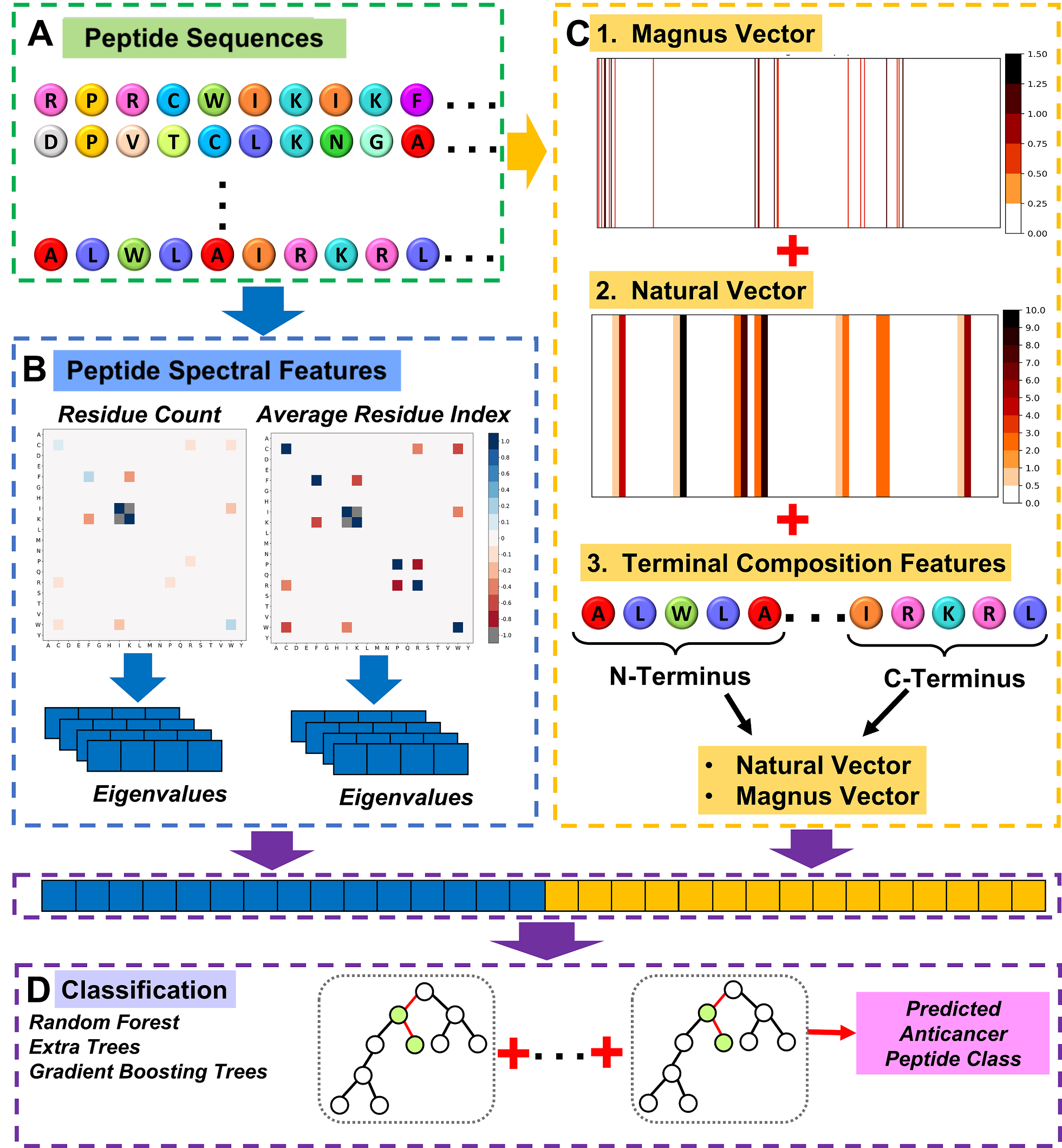}
	\caption{Pipeline of topology-enhanced machine learning model (Top-ML) for ACPs prediction. (A) The peptide sequences from AntiCP 2.0 (Dataset A or B). (B) The peptide spectral features are obtained from the sequence-based Laplacian matrices. (C) The Magnus vectors, natural vectors, and terminal composition features are generated from peptide sequences. (D) The combined features serve as inputs for the classification of ACPs and non-ACPs.}
 \label{fig:MAP-ML}
\end{figure}

In this section, we present our topology-enhanced machine learning model (Top-ML) and its results for ACPs classification. The natural vector, Magnus vector,  terminus composition features, and peptide spectral representations are combined to form the input features for our Top-ML model. Figure \ref{fig:MAP-ML} shows the general pipeline of our Top-ML model. 

\subsubsection*{Model performance on AntiCP 2.0}
We train and assess the performance of our Top-ML model using the AntiCP 2.0 benchmark datasets \cite{agrawal2021anticp}. The AntiCP 2.0 datasets comprise Dataset A and Dataset B. Dataset A consists of 970 experimentally validated ACPs and 970 randomly selected peptides, assumed to be non-ACPs. Dataset B contains 861 experimentally validated ACPs and 861 non-ACPs, which are antimicrobial peptides (AMPs). AntiCP 2.0 (Datasets A and B) was partitioned into training and test sets using an 80/20 split. We follow the same protocol as previous studies: model parameters are tuned using 5-fold cross-validation on the training data, and the performance of the Top-ML model is evaluated on the test data using various performance metrics, including accuracy, sensitivity (true positive rate), specificity (true negative rate), and the Matthews correlation coefficient (MCC). The model selection process is detailed in the Model Selection. The final Top-ML model for AntiCP 2.0 is an Extra Trees classifier with 400 trees, each using a random subset of features equal to the square root of the total number of features to determine the optimal split. In this finalized model, peptide spectral representations are derived from the mean index position, and Magnus vectors are computed using a window size of 5. The terminus used to build terminal composition features is C15 for AntiCP 2.0 (Dataset A) and N15C15 for AntiCP 2.0 (Dataset B). The final Top-ML model has been further optimized by finding the optimal probability threshold that returns the best accuracy. All reported performances of the Top-ML model are medians over 100 repeated tests.

Table \ref{tab:resultA} summarizes the results of Top-ML and existing state-of-the-art machine learning models on AntiCP 2.0 (Datasets A and B). It can be observed that on AntiCP 2.0 (Dataset A), our Top-ML model achieves one of the highest accuracies and MCC scores, outperforming several existing state-of-the-art machine learning models. This is noteworthy because these models were built on deep learning algorithms while only tree-based classifiers have been used for Top-ML. Compared to AntiCP 2.0 \cite{agrawal2021anticp}, which serves as a key benchmark model for comparison, our Top-ML model demonstrates an improvement of 1.3\% in accuracy and 3.1\% in specificity. On AntiCP 2.0 (Dataset B), the accuracy of Top-ML stands at 78.89\%, which is higher than all other models compared, except for iACP-FSCM (82.50\% accuracy) and ME-ACP(79.2\%). When compared with the benchmark model AntiCP 2.0, there is an improvement of 3.5\% in accuracy and 8.5\% in specificity. Although iACP-FSCM and ME-ACP report higher accuracy, our Top-ML model can obtain a more balanced sensitivity and specificity score. 

Additionally, our results confirm the observations made in \cite{agrawal2021anticp} that the accuracy of results on AntiCP 2.0 (Dataset A) is higher than that on AntiCP 2.0 (Dataset B). This may be attributed to the fact that ACPs exhibit greater compositional and structural similarities to AMPs, which are the negative class in AntiCP 2.0 (Dataset B), than to random peptides, which serve as the negative class in AntiCP 2.0 (Dataset A). As a result, it is observed in our numerical experiments in Model Selection that models trained on AntiCP 2.0 (Dataset A) demonstrate better performance, irrespective of the chosen model or feature encoding technique.

\begin{table}[]
	\centering
	\caption{Performance comparison of Top-ML with existing state-of-the-art machine learning models on AntiCP 2.0 (Datasets A and B). }
	\begin{tabular}{|l|cccc|}
		\hline
		 \multicolumn{5}{|c|}{AntiCP 2.0 (Dataset A)}                                                                       \\ \hline
		Model       & \multicolumn{1}{c|}{Acc}            & \multicolumn{1}{c|}{Sen}   & \multicolumn{1}{c|}{Spc}   & MCC  \\ \hline
		
		ME-ACP \cite{feng2022me}       & \multicolumn{1}{c|}{\textbf{93.30}}          & \multicolumn{1}{c|}{91.70} & \multicolumn{1}{c|}{94.80} & \textbf{0.87} \\ \hline
  \textbf{Top-ML}& \multicolumn{1}{c|}{\textbf{93.30}} & \multicolumn{1}{c|}{91.49} & \multicolumn{1}{c|}{94.85} & \textbf{0.87} \\ \hline
		iACP-DRLF \cite{lv2021anticancer}    & \multicolumn{1}{c|}{93.00}          & \multicolumn{1}{c|}{89.60} & \multicolumn{1}{c|}{\textbf{96.40}} & 0.86 \\ \hline
		AntiCP 2.0 \cite{agrawal2021anticp}   & \multicolumn{1}{c|}{92.01}          & \multicolumn{1}{c|}{\textbf{92.27}} & \multicolumn{1}{c|}{91.75} & 0.84 \\ \hline
		iACP-FSCM \cite{charoenkwan2021improved}    & \multicolumn{1}{c|}{88.90}          & \multicolumn{1}{c|}{87.60} & \multicolumn{1}{c|}{90.20} & 0.78 \\ \hline
		AntiMF \cite{liu2022antimf}       & \multicolumn{1}{c|}{91.03}          & \multicolumn{1}{c|}{88.24} & \multicolumn{1}{c|}{93.78} & 0.83 \\ \hline
		ACP-MHCNN \cite{ahmed2021acp}    & \multicolumn{1}{c|}{90.00}          & \multicolumn{1}{c|}{86.50} & \multicolumn{1}{c|}{93.30} & 0.80 \\ \hline
		AntiCP \cite{tyagi2013silico}    & \multicolumn{1}{c|}{89.95}          & \multicolumn{1}{c|}{89.69} & \multicolumn{1}{c|}{90.20} & 0.80 \\ \hline
		ACP-DL \cite{yi2019acpdl}    & \multicolumn{1}{c|}{88.10}          & \multicolumn{1}{c|}{86.00} & \multicolumn{1}{c|}{90.20} & 0.76 \\ \hline
		GRCI-Net \cite{you2021anti}   & \multicolumn{1}{c|}{87.60}          & \multicolumn{1}{c|}{87.00} & \multicolumn{1}{c|}{88.10} & 0.75 \\ \hline
		ACPred \cite{schaduangrat2019acpred}       & \multicolumn{1}{c|}{85.31}          & \multicolumn{1}{c|}{87.11} & \multicolumn{1}{c|}{83.51} & 0.71 \\ \hline
		ACPred-Fuse \cite{rao2020acpred}  & \multicolumn{1}{c|}{78.87}          & \multicolumn{1}{c|}{64.43} & \multicolumn{1}{c|}{93.30} & 0.60 \\ \hline
		iACP \cite{chen2016iacp}      & \multicolumn{1}{c|}{77.58}          & \multicolumn{1}{c|}{78.35} & \multicolumn{1}{c|}{76.80} & 0.55 \\ \hline
		PEPred-Suite \cite{wei2019pepred} & \multicolumn{1}{c|}{57.47}          & \multicolumn{1}{c|}{40.21} & \multicolumn{1}{c|}{74.74} & 0.16 \\ \hline\hline
		 \multicolumn{5}{|c|}{AntiCP 2.0 (Dataset B)}                                                                       \\ \hline
		Models       & \multicolumn{1}{c|}{Acc}   & \multicolumn{1}{c|}{Sen}    & \multicolumn{1}{c|}{Spc}   & MCC  \\ \hline
		iACP-FSCM \cite{charoenkwan2021improved}    & \multicolumn{1}{c|}{\textbf{82.50}} & \multicolumn{1}{c|}{72.60}  & \multicolumn{1}{c|}{\textbf{90.30}} & \textbf{0.65} \\ \hline
		
		ME-ACP \cite{feng2022me}        & \multicolumn{1}{c|}{79.20} & \multicolumn{1}{c|}{74.90}  & \multicolumn{1}{c|}{81.18} & 0.58 \\ \hline
    \textbf{Top-ML}& \multicolumn{1}{c|}{78.89} & \multicolumn{1}{c|}{76.90}  & \multicolumn{1}{c|}{81.96} & 0.63 \\ \hline
		iACP-DRLF \cite{lv2021anticancer}     & \multicolumn{1}{c|}{77.50} & \multicolumn{1}{c|}{80.70}  & \multicolumn{1}{c|}{74.30} & 0.55 \\ \hline
		AntiMF \cite{liu2022antimf}       & \multicolumn{1}{c|}{75.94} & \multicolumn{1}{c|}{65.34}  & \multicolumn{1}{c|}{86.63} & 0.53 \\ \hline
		AntiCP 2.0 \cite{agrawal2021anticp}   & \multicolumn{1}{c|}{75.43} & \multicolumn{1}{c|}{77.46}  & \multicolumn{1}{c|}{73.41} & 0.51 \\ \hline
		ACP-DL \cite{yi2019acpdl}     & \multicolumn{1}{c|}{74.60} & \multicolumn{1}{c|}{74.90}  & \multicolumn{1}{c|}{74.30} & 0.49 \\ \hline
		GRCI-Net \cite{you2021anti}   & \multicolumn{1}{c|}{74.60} & \multicolumn{1}{c|}{75.40}  & \multicolumn{1}{c|}{73.70} & 0.49 \\ \hline
		ACP-MHCNN \cite{ahmed2021acp}    & \multicolumn{1}{c|}{74.00} & \multicolumn{1}{c|}{73.70}  & \multicolumn{1}{c|}{74.30} & 0.48 \\ \hline
		ACPred-Fuse \cite{rao2020acpred}  & \multicolumn{1}{c|}{68.90} & \multicolumn{1}{c|}{69.19}  & \multicolumn{1}{c|}{68.60} & 0.38 \\ \hline
		iACP  \cite{chen2016iacp}       & \multicolumn{1}{c|}{55.10} & \multicolumn{1}{c|}{77.91}  & \multicolumn{1}{c|}{32.16} & 0.11 \\ \hline
		PEPred-Suite \cite{wei2019pepred} & \multicolumn{1}{c|}{53.49} & \multicolumn{1}{c|}{33.14}  & \multicolumn{1}{c|}{73.84} & 0.08 \\ \hline
		ACPred  \cite{schaduangrat2019acpred}     & \multicolumn{1}{c|}{53.47} & \multicolumn{1}{c|}{85.55}  & \multicolumn{1}{c|}{21.39} & 0.09 \\ \hline
		AntiCP \cite{tyagi2013silico}      & \multicolumn{1}{c|}{50.58} & \multicolumn{1}{c|}{\textbf{100.00}} & \multicolumn{1}{c|}{01.16} & 0.07 \\ \hline
	\end{tabular}
 \label{tab:resultA}
\end{table}

\subsubsection*{Model performance on mACPpred2.0}
\begin{figure}[ht]

	\centering
	\includegraphics[width=\linewidth]{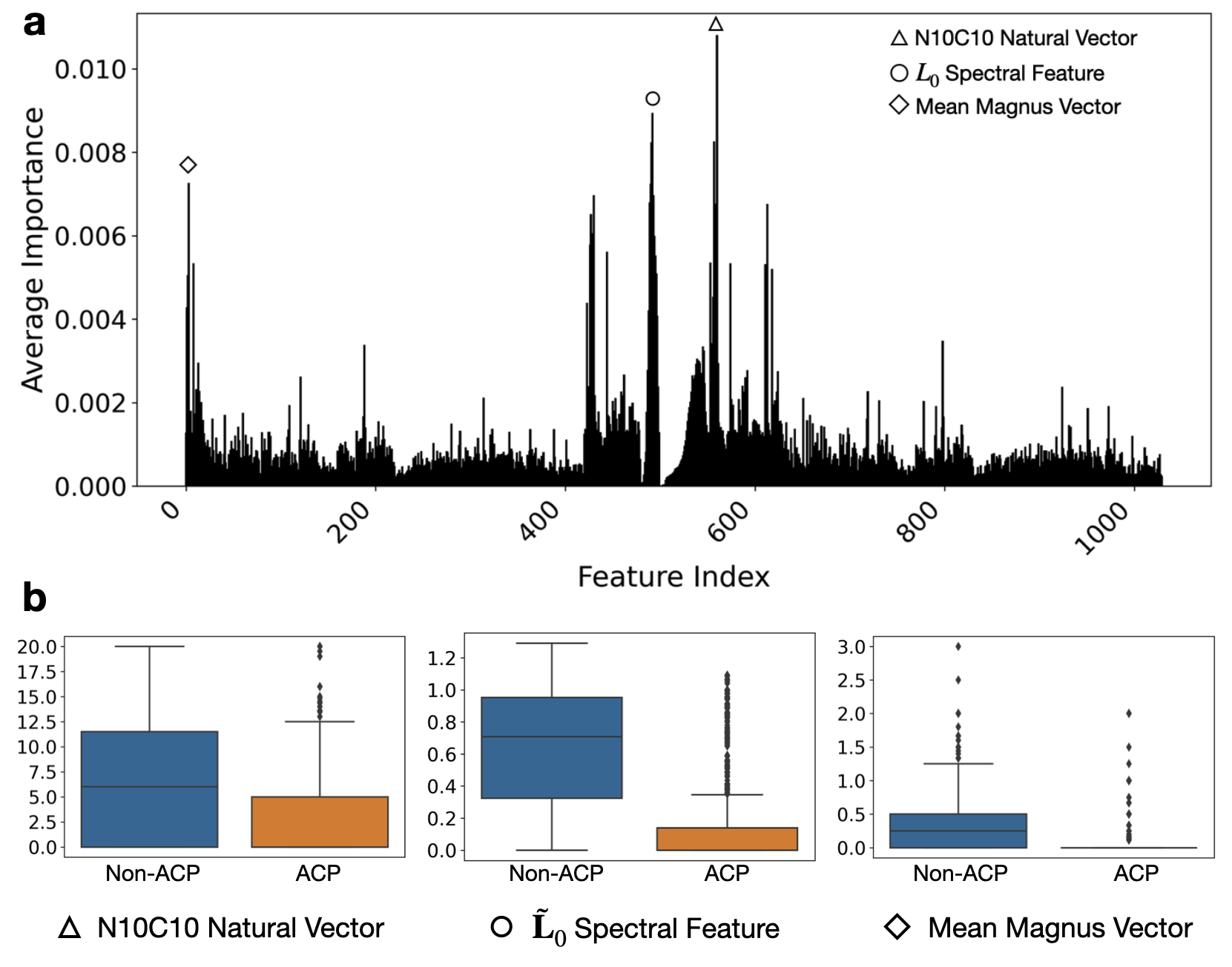}
	\caption{Feature importance of the Top-ML model, averaged over 100 iterations of training using the mACPpred 2.0 training set. Feature importance is computed using the Gini importance criterion. We identified the three highest peaks in the top plot, highlighted with markers, and plotted their value distributions for each class in the bottom plot.}
	\label{fig:importance}
\end{figure}

To demonstrate the generalizability of our method, we tested the model on a more recent and comprehensive ACP dataset, mACPpred 2.0 \cite{sangaraju2024macppred}. This dataset was compiled from 11 existing studies, excluding amino acid sequences shorter than 5 or longer than 50 residues. It also accounts for sequence similarities by removing redundancy using the CD-HIT technique with a similarity threshold of 0.85. The training dataset comprises 1,176 ACPs and 1,176 non-ACPs, while the test dataset consists of 610 ACPs and 2,760 non-ACPs. We selected the N10C10 terminus to construct terminal composition features for this dataset, and the performance of our model compared to other benchmark models is presented in Table \ref{tab:mACPpred2} below. Our tree-based model achieves similar accuracy when compared to other more complex deep learning models such as mACPpred 2.0 \cite{sangaraju2024macppred} and MLACP 2.0 \cite{park2022mlacp}. Both of these models use complex deep learning framework that integrate probabilistic features computed from various baseline models. \begin{revised}
For example, mACPpred2.0 \cite{boopathi2019macppred} uses a stacked deep learning approach, combining multiple base learners trained on different feature types with a 1D CNN that integrates the predictions of base models into the final output. Its features also include 16 pre-trained NLP-based embeddings, leveraging external datasets. In contrast, our model employs vectors directly derived from amino acid sequences and uses a tree-based classifier, offering better computational efficiency and greater interpretability, as demonstrated by the feature importance analysis in Figure \ref{fig:importance}.
\end{revised}

\begin{table}[H]
	\centering
	\caption{Performance comparison of Top-ML with existing state-of-the-art machine learning models on mACPpred 2.0. }
	\begin{tabular}{|l|c|c|c|c|}
		\hline                                                        		
        Model       & \multicolumn{1}{c|}{Acc}            & \multicolumn{1}{c|}{Sen}   & \multicolumn{1}{c|}{Spc}   & MCC  \\ \hline
		
        mACPpred  2.0 \cite{sangaraju2024macppred} & \textbf{85.7} & \textbf{80.7} & 86.8 & \textbf{59.7} \\ \hline
        MLACP 2.0 \cite{park2022mlacp} & 83.8 & 79.2 & 84.9 & 55.7 \\ \hline
        AntiCP 2.0 (Alternate) \cite{agrawal2021anticp} & 83.6 & 60.5 & 88.7 & 47.2 \\ \hline
        \textbf{Top-ML}  & 82.5 & 74.4 & 84.3 & 51.4 \\ \hline
        iDACP (Sp of 90\%) \cite{huang2021identification} & 82.5 & 20.3 & 96.3 & 25.5 \\ \hline
        iDACP (Sp of 80\%) \cite{huang2021identification} & 82.4 & 35.6 & 92.7 & 33.0 \\ \hline
        iDACP (Sp of 70\%) \cite{huang2021identification} & 81.8 & 48.4 & 89.2 & 38.0 \\ \hline
        PreTP-2L \cite{yan2023pretp} & 81.8 & 2.1 & \textbf{99.4} & 6.5 \\ \hline
        iDACP (Sp of 60\%) \cite{huang2021identification} & 81.7 & 61.6 & 86.1 & 44.0 \\ \hline
        CancerGram \cite{burdukiewicz2020cancergram} & 81.5 & 6.6 & 98.1 & 11.0 \\ \hline
        iAMP-RAAC \cite{dong2021amino} & 81.4 & 6.6 & 97.9 & 10.2 \\ \hline
        mACPpred \cite{boopathi2019macppred} & 80.7 & 75.4 & 81.9 & 48.9 \\ \hline
        iDACP (Sp of 50\%) \cite{huang2021identification} & 80.5 & 68.0 & 83.3 & 45.1 \\ \hline
        AMPfun \cite{chung2020characterization} & 76.0 & 76.9 & 75.8 & 42.9 \\ \hline
        ACPred-BMF (Alternate) \cite{han2022acpred} & 75.9 & 44.1 & 82.9 & 25.1 \\ \hline
        ACPred \cite{schaduangrat2019acpred} & 75.1 & 51.0 & 80.4 & 27.8 \\ \hline
        TriNet \cite{zhou2023trinet} & 74.7 & 43.3 & 81.7 & 22.9 \\ \hline
        ACP-MHCNN (740) \cite{ahmed2021acp} & 71.1 & 35.7 & 78.9 & 13.2 \\ \hline
        AntiCP 2.0 (Main) \cite{agrawal2021anticp} & 71.0 & 43.3 & 77.2 & 17.8 \\ \hline
        ACP-MHCNN (500/164) \cite{ahmed2021acp} & 61.6 & 16.7 & 71.6 & -10.2 \\ \hline
        ACPred-BMF (Main) \cite{han2022acpred} & 60.0 & 67.2 & 58.4 & 19.8 \\ \hline
        iAMPCN \cite{xu2023iampcn} & 24.1 & 62.5 & 15.6 & -21.3 \\ \hline
		 \end{tabular}
 \label{tab:mACPpred2}
\end{table}

In Figure \ref{fig:importance}a, we present the feature importance calculated from the Top-ML model using the Gini importance criterion, which measures the total reduction in Gini impurity contributed by each feature. To understand how these feature values vary between non-ACPs and ACPs, we identified the three highest peaks in Figure \ref{fig:importance}a and plotted the distribution of the feature values for each class in Figure \ref{fig:importance}b. The largest spike corresponds to a feature extracted from the natural vector constructed from the N10C10 terminus vector, representing the mean position of the amino acid ``E'' in the N10C10 terminus. On average, the amino acid ``E'', or glutamic acid, appears later in non-ACPs than in ACPs. The second-highest spike corresponds to a spectrum from the Hodge Laplacian $\tilde{\mathbf{L}}_0$, where non-ACPs exhibit a larger spectrum than ACPs. The spectrum of $\tilde{\mathbf{L}}_0$ is associated with clustering patterns \cite{von2007tutorial}, where smaller values indicate that amino acids from the same clusters are more likely to appear adjacent to each other. \begin{revised}
The plot suggests that ACPs exhibit a stronger clustering pattern than non-ACPs.\end{revised}
 The rightmost figure displays a feature corresponding to the third-highest peak, derived from the mean Magnus vector, which represents the occurrence of glutamic acid. The plot suggests that glutamic acid is more frequent in non-ACPs than in ACPs. \begin{revised}
This observation is consistent with the analyses presented in \cite{huang2021identification, tyagi2013silico}. This could be explained by the fact that ACPs tend to have high hydrophobicity and a positive net charge to effectively interact with negatively charged cancer cells \cite{huang2021identification}, whereas glutamic acid is negatively charged and hydrophilic. 
\end{revised}

\section*{Discussion}
In this paper, we propose a topology-enhanced machine learning model (Top-ML) that combines peptide spectral features with Magnus and natural vector-based peptide features. We demonstrate that our Top-ML model based on an Extra Trees classifier outperforms or achieves similar level of accuracy as several existing state-of-the-art deep learning models for ACPs classification on benchmark datasets AntiCP 2.0 and mACPpred 2.0. Our results highlight the potential of mathematics-assisted peptide-based features to improve the performance of ACPs identification. Furthermore, our model provides improved interpretability, demonstrated through feature importance analysis on the mACPpred 2.0 dataset.

 Since this study focuses only on the topological featurization of peptides, it has not considered physicochemical properties such as hydrophobicity and helicity, which have been shown to also provide valuable signals in encoding a peptide. \begin{revised} Incorporating the 3D structures, for instance through spectral features computed using the simplicial complex representation, could be an interesting direction for future research. \end{revised} Our framework focus on developing expressive feature extraction methods for sequential data. Therefore, while our experiments focused on anticancer peptides due to data availability and their biomedical significance, the framework is inherently applicable to various sequence types, including amino acid (proteins and peptides) and nucleotide (DNA and RNA) sequences. Future research could investigate its generalizability and effectiveness across other datasets and various prediction or regression tasks. For instance, most existing peptide datasets consist of natural peptides. Expanding the framework to include modified peptides could offer valuable insights into their potential in drug application.
\section*{Method}

\subsection*{Combinatorial Laplacian}
Our sequence-based Laplacian construction is heavily inspired by combinatorial Laplacian matrices constructed from simplicial complexes. Here, we provide the background information on combinatorial Laplacian matrices. A simplicial complex $K$ is made of simplexes. A $p$-simplex is a geometric object formed by $p+1$ affinely independent points, resulting in a convex hull. For example, a $0$-simplex is a node, a $1$-simplex is an edge, a $2$-simplex is a triangle (with a solid inside region), and a $3$-simplex is a solid tetrahedron. Mathematically, a $p$-simplex $\sigma^p = \{v_0, v_1, v_2, \cdots, v_p\}$ can be written as follows:

\[
\sigma^p = \bigg\{\lambda_0v_0+\lambda_1v_1+\cdots+\lambda_pv_p \bigg|\sum_{i=0}^p \lambda_i = 0;\forall i, 0\leq \lambda_i \leq 1\bigg\}.
\]

In order to construct combinatorial Laplacian matrices from a simplicial complex $K$, the boundary matrices are first computed by obtaining the orientation and relevant faces for each simplex in $K$. Here, $\sigma_i^{k-1} \subset \sigma_j^k$ indicates that $\sigma_i^{k-1}$ is a \textit{face} of $\sigma_j^k$. In this case, $\sigma_j^k$ is called a \textit{co-face} of $\sigma_i^{k-1}$. Otherwise, we denote $\sigma_i^{k-1} \not\subset \sigma_j^k$. Additionally, $\sigma_i^{k-1} \sim \sigma_j^k$ indicates that the two simplices are similarly oriented and $\sigma_i^{k-1} \not\sim \sigma_j^k$ means two simplices are dissimilarly oriented. For an oriented simplicial complex, its $k$-th boundary matrix $\mathbf{B}_k$ can be defined as follows:

\begin{equation}
\label{eqn:boundary}
\mathbf{B}_k(i,j) = \begin{cases}
1, & \text{if } \sigma_i^{k-1}\subset \sigma_j^k \text{ and } \sigma_i^{k-1}\sim \sigma_j^k \\
-1, & \text{if } \sigma_i^{k-1}\subset \sigma_j^k \text{ and } \sigma_i^{k-1}\nsim \sigma_j^k\\
0, & \text{if } \sigma_i^{k-1}\not\subset \sigma_j^k.
\end{cases}
\end{equation}

Laplacian matrices encapsulate the topological and geometrical information of the simplicial complex $K$. Once the boundary matrices are obtained, the $k$-th combinatorial Laplacian matrices can be computed as follows:
\begin{equation}
\mathbf{L}_k = \begin{cases}
\mathbf{B}_1\mathbf{B}_1^T, & \text{if } k=0 \\
\mathbf{B}_k^T\mathbf{B}_k + \mathbf{B}_{k+1}\mathbf{B}_{k+1}^T, & \text{if } k \geq 1.
\end{cases}
\label{eq:laplacian}
\end{equation}

\begin{figure}[ht]
	\centering
	\includegraphics[width=\linewidth]{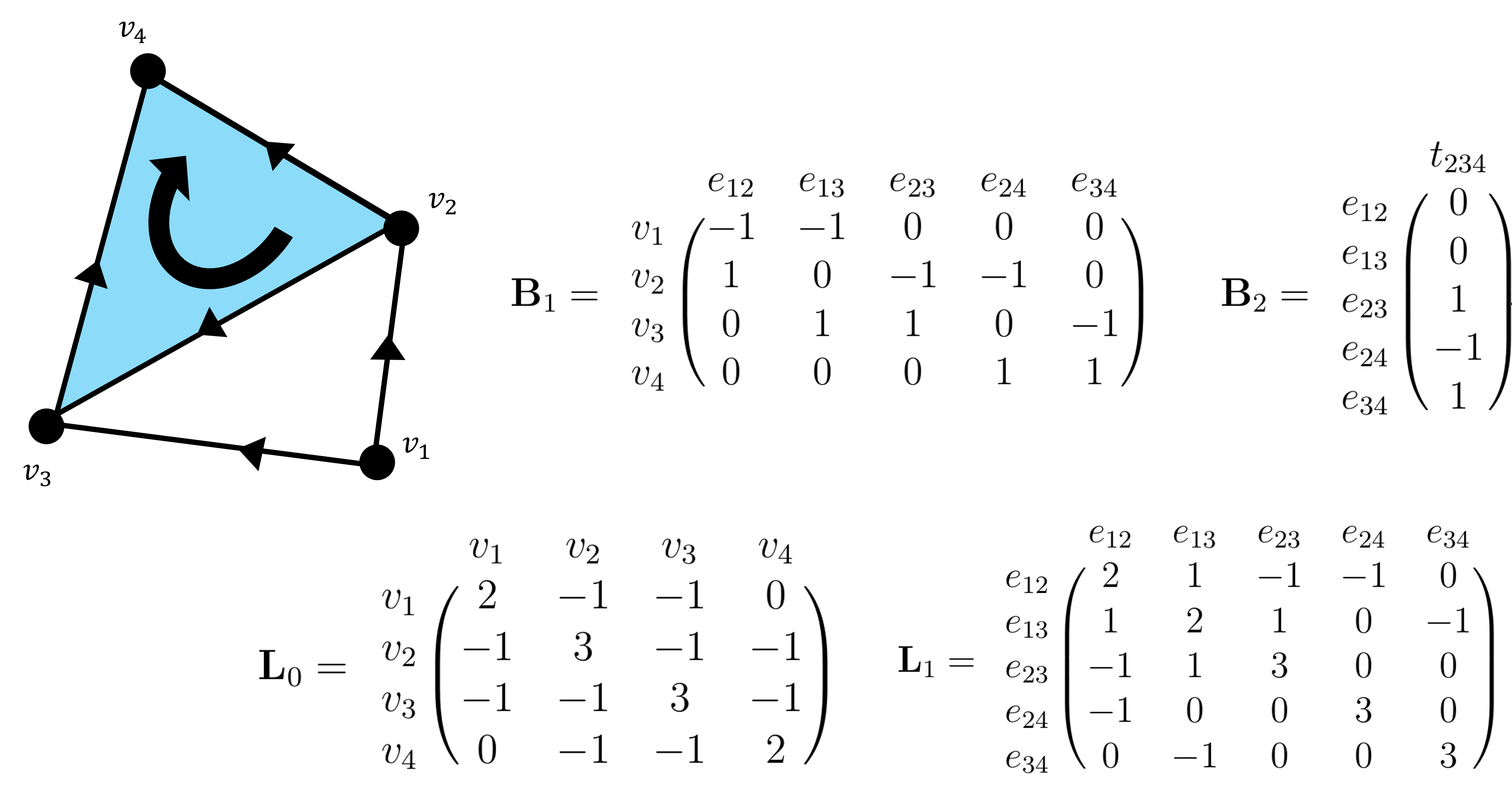}
	\caption{Illustration of boundary and combinatorial Laplacian matrices for an oriented simplicial complex.}
	\label{fig:example}
\end{figure}

For a simple illustration, refer to Figure \ref{fig:example}, which displays an oriented simplicial complex $K$  along with its corresponding boundary and Laplacian matrices. For the combinatorial Laplacian matrices $\mathbf{L}_0$ and $\mathbf{L}_1$, every diagonal entry corresponds to the degree of a $k$-simplex $\sigma$ (i.e., the total number of faces and co-faces of $\sigma$).  The symmetric normalized Laplacian is defined as follows:
\begin{equation}
    \tilde{\mathbf{L}}_k = \mathbf{D}^\frac{1}{2}_k\mathbf{L}_k\mathbf{D}_k^{-\frac{1}{2}} \label{eq:normalized L},
\end{equation}
where $\mathbf{D}_k$ is a $k$-th diagonal matrix with diagonal entries from $\mathbf{L}_k$.

\subsection*{Model Selection}
Three classifiers were considered to train our Top-ML features: Random Forest (RF), Extra Trees (Etrees), and Gradient Boosting Trees (GBT). The set of features tested consists of the natural vector, mean and sum Magnus vector , while the $L_0$ and $L_1$ peptide spectral features have been aggregated using mean and sum. The set of features was then augmented with additional feature classes generated using the terminus composition representation. 

The hyperparameters for both the Random Forest and Extra Trees classifiers have been configured as follows during experimentation: a total of 400 trees are used, with each tree considering a subset of features equal to the square root of the total number of features. The splitting process is evaluated using the Gini impurity. For the Gradient Boosting classifier, the hyperparameters are configured as follows: 400 boosting stages, a learning rate of 0.01, a maximum depth of 8, and a subsample ratio of 0.8. Furthermore, each combination of feature inputs has been systematically studied. The results shown for each model are the medians over 100 iterations. The threshold-dependent metrics, including accuracy, specificity, sensitivity, and the Matthews correlation coefficient, are generated at the 0.5 probability threshold.

\subsubsection*{Peptide Spectral Features}
In the following experiment, we systematically tested two definitions of $f(w)$ in Equation (\ref{eqn:Bk}) for constructing Laplacian matrices: the number of occurrences of amino acid $w$ (frequency), and its mean index position. \begin{revised}
We compared the 5-fold cross-validation performance on the training data of two construction methods using three types of classifiers     
\end{revised}: Random Forest (RF), Extra Trees (ETrees), and Gradient Boosting Trees (GBT). Models are evaluated on the benchmark dataset AntiCP 2.0, using peptide spectral features derived from $L_0$ and $L_1$ as inputs. 

As shown in Table \ref{tbl:SpectPepA}, on AntiCP 2.0 (Dataset A), when controlling for the classifier, the mean index position definition achieves higher accuracy and AUROC compared to the frequency. GBT with mean index position-based features has the highest accuracy of 80.16\%. On AntiCP 2.0 (Dataset B), the mean position-based feature outperforms the frequency-based feature when using Etrees, whereas the frequency-based spectral feature performs better with GBT and RF. The highest accuracy of 71.99\% is achieved using Random Forest when peptide spectral features are constructed from the frequency. For the final model, we selected the mean index position feature because it outperforms the frequency feature in more settings. Consequently, we chose Etrees as the classifier due to its higher accuracy with mean index position-based features.

\begin{table}[ht]
	\centering
        %\small  % or \footnotesize for even smaller text

	\begin{tabular}{|l|l|c|c|c|c|c|}
		\hline
             \multicolumn{7}{|c|}{AntiCP 2.0 (Dataset A)}\\ \hline
		Spectral Feature             & Classifier    & Acc            & Sen   & Spc   & MCC  & AUROC \\ \hline
		\multirow{3}{4em}{Mean Position} & Etrees & 79.64 & 76.80 & 82.47 & 0.59 & 0.80\\ 
		%\multicolumn{1}{|l|}{Mean Position}
            & RF & 79.64 & 77.32 & 81.96 & 0.59 & 0.80\\ 
            & GBT & 80.16  & 78.35 & 81.96 & 0.60 & 0.80\\ \hline
		\multirow{3}{4em}{Frequency} & Etrees & 79.38 & 77.84 & 80.93 & 0.59 & 0.79\\ 
		& RF & 78.35 & 77.06 & 79.90 & 0.57 & 0.78\\ 
            & GBT& 78.87 & 77.84 & 79.90 & 0.58 & 0.79\\ \hline\hline
             \multicolumn{7}{|c|}{AntiCP 2.0 (Dataset B)}\\ \hline
		\multicolumn{1}{|l|}{Spectral Feature}              & Classifier    & \multicolumn{1}{c|}{Acc}            & \multicolumn{1}{c|}{Sen}   & \multicolumn{1}{c|}{Spc}   & \multicolumn{1}{c|}{MCC}  & AUROC \\ \hline
        \multirow{3}{4em}{Mean Position} & Etrees   & 70.67 & 67.25 & 74.12 & 0.41 & 0.71\\ 
	& RF & 70.38 & 66.67 & 74.12 & 0.41 & 0.70\\ 
		& GBT & 68.62 & 64.91 & 72.35 & 0.37 & 0.69\\ \hline
		\multirow{3}{4em}{Frequency} & Etrees& 70.38 & 63.74 & 76.47 & 0.41 & 0.70\\ 
		& RF & 71.99 & 63.74 & 80.00 & 0.45 & 0.72\\ 
		& GBT & 70.67 & 63.16 & 78.24 & 0.42 & 0.71\\ \hline
	\end{tabular}
\caption{The 5-fold cross-validation performance using peptide spectral features on AntiCP 2.0 (Datasets A and B) }
 \label{tbl:SpectPepA}
\end{table}

\subsubsection*{Sequence-based Features}
\paragraph{Magnus Vector}
In this 5-fold cross-validation experiment on the training data, detailed in Table \ref{tbl:MagnusA}, we focus on the Extra Trees classifier and evaluate its performance using the mean Magnus vector across various window sizes. A window size of 4 yields the highest accuracy on AntiCP 2.0 (Dataset A), whereas a window size of 5 achieves the highest accuracy on AntiCP 2.0 (Dataset B). We selected a window size of 5 for our final Top-ML model because it achieves the highest AUROC for both AntiCP 2.0 Datasets A and B, the highest accuracy of 76.54\% on AntiCP 2.0 Dataset B, and the second-highest accuracy of 92.17\% on AntiCP 2.0 Dataset A.

\begin{table}[H]
	\centering
         %\small  % or \footnotesize for even smaller text

	\begin{tabular}{|ll|ccccc|}
		\hline
		\multicolumn{7}{|c|}{AntiCP 2.0 (Dataset A)}                                                                                                    \\ \hline
		\multicolumn{1}{|l|}{Window Size} & Classifier  & \multicolumn{1}{c|}{Acc}            & \multicolumn{1}{c|}{Sen}   & \multicolumn{1}{c|}{Spc}   & \multicolumn{1}{c|}{MCC}  & AUROC \\ \hline
		\multicolumn{1}{|l|}{4}           & Etrees & \multicolumn{1}{c|}{\textbf{92.53}}          & \multicolumn{1}{c|}{88.56} & \multicolumn{1}{c|}{\textbf{95.77}} & \multicolumn{1}{c|}{\textbf{0.85}} & \textbf{0.92}\\ \hline
		\multicolumn{1}{|l|}{5}           & Etrees & \multicolumn{1}{c|}{92.17} & \multicolumn{1}{c|}{88.66} & \multicolumn{1}{c|}{95.36} & \multicolumn{1}{c|}{0.84} & \textbf{0.92}\\ \hline
		\multicolumn{1}{|l|}{6}           & Etrees & \multicolumn{1}{c|}{91.69}          & \multicolumn{1}{c|}{\textbf{90.36}} & \multicolumn{1}{c|}{93.08} & \multicolumn{1}{c|}{0.83} & \textbf{0.92}\\ \hline
		\multicolumn{1}{|l|}{7}           & Etrees & \multicolumn{1}{c|}{91.80}          & \multicolumn{1}{c|}{89.64} & \multicolumn{1}{c|}{94.05} & \multicolumn{1}{c|}{0.84} & \textbf{0.92}\\ \hline\hline
        \multicolumn{7}{|c|}{AntiCP 2.0 (Dataset B)}                                                                                                    \\ \hline
		\multicolumn{1}{|l|}{Window Size} & Classifier  & \multicolumn{1}{c|}{Acc}            & \multicolumn{1}{c|}{Sen}   & \multicolumn{1}{c|}{Spc}   & \multicolumn{1}{c|}{MCC}  & AUROC \\ \hline
		\multicolumn{1}{|l|}{4}           & Etrees & \multicolumn{1}{c|}{75.95}          & \multicolumn{1}{c|}{\textbf{74.85}} & \multicolumn{1}{c|}{77.06} & \multicolumn{1}{c|}{0.52} & 0.76\\ \hline
		\multicolumn{1}{|l|}{5}           & Etrees & \multicolumn{1}{c|}{\textbf{76.54}} & \multicolumn{1}{c|}{74.27} & \multicolumn{1}{c|}{\textbf{78.82}} & \multicolumn{1}{c|}{\textbf{0.53}} & \textbf{0.77}\\ \hline
		\multicolumn{1}{|l|}{6}           & Etrees & \multicolumn{1}{c|}{75.52}          & \multicolumn{1}{c|}{72.78} & \multicolumn{1}{c|}{78.24} & \multicolumn{1}{c|}{0.51} & 0.76\\ \hline
		\multicolumn{1}{|l|}{7}           & Etrees & \multicolumn{1}{c|}{75.67}          & \multicolumn{1}{c|}{72.78} & \multicolumn{1}{c|}{78.57} & \multicolumn{1}{c|}{0.52} & 0.76\\ \hline
	\end{tabular}
 \caption{The 5-fold cross-validation performance using Magnus vector representation on AntiCP 2.0 (Datasets A and B)}
 \label{tbl:MagnusA}
\end{table}

\paragraph{Natural Vector}
In this section, we tested three classifiers, i.e., RF, ETrees, and GBT, using the natural vector as input features \begin{revised}
with 5-fold cross-validation on the training data. \end{revised}The Gradient Boosting Trees classifier provides the best accuracy of 92.53\% on AntiCP 2.0 (Dataset A) (Table \ref{tbl:naturalA}). On AntiCP 2.0 (Dataset B), Extra Trees achieves the highest accuracy of 74.78\%. For both Magnus representation and natural representation, the specificity metric generally trends higher than sensitivity across different classifiers.

\begin{table}[H]
	\centering
	\caption{The 5-fold cross-validation performance using natural vector-based features on AntiCP 2.0 (Datasets A and B)}
        %\small  % or \footnotesize for even smaller text
	
 \begin{tabular}{|l|l|ccccc|}
		\hline
        \multicolumn{7}{|c|}{AntiCP 2.0 (Dataset A)}                                                                                                                                    \\ \hline
		Features       & Classifier    & \multicolumn{1}{c|}{Acc}            & \multicolumn{1}{c|}{Sen}   & \multicolumn{1}{c|}{Spc}   & \multicolumn{1}{c|}{MCC}  & \multicolumn{1}{l|}{AUROC} \\ \hline
		Natural Vector & Etrees& \multicolumn{1}{c|}{92.40} & \multicolumn{1}{c|}{88.14} & \multicolumn{1}{c|}{\textbf{96.91}} & \multicolumn{1}{c|}{\textbf{0.85}} & 0.92\\ \hline
		Natural Vector & RF& \multicolumn{1}{c|}{92.27}          & \multicolumn{1}{c|}{89.69} & \multicolumn{1}{c|}{94.85} & \multicolumn{1}{c|}{\textbf{0.85}} & 0.92\\ \hline
		Natural Vector & GBT& \multicolumn{1}{c|}{\textbf{92.53}}          & \multicolumn{1}{c|}{\textbf{90.21}} & \multicolumn{1}{c|}{94.85} & \multicolumn{1}{c|}{\textbf{0.85}} & \textbf{0.93}\\ \hline\hline
        \multicolumn{7}{|c|}{AntiCP 2.0 (Dataset B)}                                                                                                                         \\ \hline
		Features       & Classifier    & \multicolumn{1}{c|}{Acc}            & \multicolumn{1}{c|}{Sen}   & \multicolumn{1}{c|}{Spc}   & \multicolumn{1}{c|}{MCC}  & \multicolumn{1}{l|}{AUROC} \\ \hline
		Natural Vector & Etrees& \multicolumn{1}{c|}{\textbf{74.78}} & \multicolumn{1}{c|}{73.10} & \multicolumn{1}{c|}{\textbf{76.47}} & \multicolumn{1}{c|}{\textbf{0.50}} & \textbf{0.75}\\ \hline
		Natural Vector & RF& \multicolumn{1}{c|}{74.19}          & \multicolumn{1}{c|}{73.10} & \multicolumn{1}{c|}{75.29} & \multicolumn{1}{c|}{0.48} & 0.74                       \\ \hline
		Natural Vector & GBT& \multicolumn{1}{c|}{74.19}          & \multicolumn{1}{c|}{\textbf{74.27}} & \multicolumn{1}{c|}{74.12} & \multicolumn{1}{c|}{0.48} & 0.74
\\ \hline
	\end{tabular}
 \label{tbl:naturalA}
\end{table}

\paragraph{Terminal Composition Features}
To test which terminus to use, we generate the natural vector and mean Magnus vector based on the terminus instead of the entire peptide sequence and test the performance using the Extra Trees classifier \begin{revised} with 5-fold cross validation on the training data. \end{revised} For AntiCP 2.0 (Dataset A) (Table \ref{tbl:terminalA}), the C15-based natural and Magnus vectors outperformed all other terminal composition encoding with an accuracy score of 92.78\%. On AntiCP 2.0 (Dataset B) shown in Table \ref{tbl:terminalA}, the N15C15 terminus composition feature provided the best results with an accuracy of 77.13\%.

Considering the different feature encodings individually using the Extra Trees classifier, it has been observed that the Magnus representation and natural vector representation yield better results than the terminal composition feature and the peptide spectral representation. The natural vector and Magnus vector have similar accuracy on AntiCP 2.0 (Dataset A), while the Magnus representation performs better on AntiCP 2.0 (Dataset B).

\begin{table}[H]
	\centering
	\caption{The 5-fold cross-validation performance using Terminal composition features on AntiCP 2.0 (Datasets A and B)}
	        \small  % or \footnotesize for even smaller text
        \begin{tabular}{|l|l|ccccc|}
		\hline
		\multicolumn{7}{|c|}{AntiCP 2.0 (Dataset A)}                                                                                                                         \\ \hline
		Features & Classifier  & \multicolumn{1}{c|}{Acc}            & \multicolumn{1}{c|}{Sen}   & \multicolumn{1}{c|}{Spc}   & \multicolumn{1}{c|}{MCC}  & \multicolumn{1}{l|}{AUROC} \\ \hline
		N5       & Etrees & \multicolumn{1}{c|}{85.83}          & \multicolumn{1}{c|}{84.02} & \multicolumn{1}{c|}{87.63} & \multicolumn{1}{c|}{0.72} & 0.86\\ \hline
		C5       & Etrees & \multicolumn{1}{c|}{84.02}          & \multicolumn{1}{c|}{79.38} & \multicolumn{1}{c|}{88.66} & \multicolumn{1}{c|}{0.68} & 0.84\\ \hline
		N5C5     & Etrees & \multicolumn{1}{c|}{89.05}          & \multicolumn{1}{c|}{85.57} & \multicolumn{1}{c|}{92.78} & \multicolumn{1}{c|}{0.78} & 0.89\\ \hline
		N10      & Etrees & \multicolumn{1}{c|}{89.43}          & \multicolumn{1}{c|}{85.05} & \multicolumn{1}{c|}{93.81} & \multicolumn{1}{c|}{0.79} & 0.89
\\ \hline
		C10      & Etrees & \multicolumn{1}{c|}{87.1}          & \multicolumn{1}{c|}{79.90} & \multicolumn{1}{c|}{94.33} & \multicolumn{1}{c|}{0.75} & 0.87\\ \hline
		N10C10   & Etrees & \multicolumn{1}{c|}{90.46}          & \multicolumn{1}{c|}{85.05} & \multicolumn{1}{c|}{95.88} & \multicolumn{1}{c|}{0.81} & 0.91\\ \hline
		N15      & Etrees & \multicolumn{1}{c|}{89.95}          & \multicolumn{1}{c|}{82.99} & \multicolumn{1}{c|}{\textbf{96.91}} & \multicolumn{1}{c|}{0.81} & 0.90\\ \hline
		C15      & Etrees & \multicolumn{1}{c|}{\textbf{92.78}} & \multicolumn{1}{c|}{\textbf{89.69}} & \multicolumn{1}{c|}{95.88} & \multicolumn{1}{c|}{\textbf{0.86}} & \textbf{0.93}\\ \hline
		N15C15   & Etrees & \multicolumn{1}{c|}{91.50}          & \multicolumn{1}{c|}{86.60} & \multicolumn{1}{c|}{95.88} & \multicolumn{1}{c|}{0.83} & 0.92\\ \hline\hline
        \multicolumn{7}{|c|}{AntiCP 2.0 (Dataset B)}                                                                                                                         \\ \hline
		Features & Classifier  & \multicolumn{1}{c|}{Acc}            & \multicolumn{1}{c|}{Sen}   & \multicolumn{1}{c|}{Spc}   & \multicolumn{1}{c|}{MCC}  & \multicolumn{1}{l|}{AUROC} \\ \hline
		N5       & Etrees & \multicolumn{1}{c|}{75.37}          & \multicolumn{1}{c|}{\textbf{75.44}} & \multicolumn{1}{c|}{75.29} & \multicolumn{1}{c|}{0.51} & 0.75\\ \hline
		C5       & Etrees & \multicolumn{1}{c|}{67.30}          & \multicolumn{1}{c|}{67.25
} & \multicolumn{1}{c|}{67.65
} & \multicolumn{1}{c|}{0.35
} & 0.67
\\ \hline
		N5C5     & Etrees & \multicolumn{1}{c|}{75.07}          & \multicolumn{1}{c|}{71.93} & \multicolumn{1}{c|}{78.24} & \multicolumn{1}{c|}{0.50} & 0.75\\ \hline
		N10      & Etrees & \multicolumn{1}{c|}{75.95}          & \multicolumn{1}{c|}{74.27} & \multicolumn{1}{c|}{78.24} & \multicolumn{1}{c|}{0.52} & 0.76\\ \hline
		C10      & Etrees & \multicolumn{1}{c|}{73.61}          & \multicolumn{1}{c|}{70.18} & \multicolumn{1}{c|}{77.06} & \multicolumn{1}{c|}{0.47} & 0.74\\ \hline
		N10C10   & Etrees & \multicolumn{1}{c|}{75.66}          & \multicolumn{1}{c|}{74.27} & \multicolumn{1}{c|}{77.06} & \multicolumn{1}{c|}{0.51} & 0.76\\ \hline
		N15      & Etrees & \multicolumn{1}{c|}{75.66}          & \multicolumn{1}{c|}{73.10} & \multicolumn{1}{c|}{78.24} & \multicolumn{1}{c|}{0.51} & 0.76\\ \hline
		C15      & Etrees & \multicolumn{1}{c|}{76.69}          & \multicolumn{1}{c|}{74.85} & \multicolumn{1}{c|}{78.82} & \multicolumn{1}{c|}{0.53} & \textbf{0.77}\\ \hline
		N15C15   & Etrees & \multicolumn{1}{c|}{\textbf{77.13}} & \multicolumn{1}{c|}{74.85} & \multicolumn{1}{c|}{\textbf{79.41}} & \multicolumn{1}{c|}{\textbf{0.54}} & \textbf{0.77}\\ \hline
	\end{tabular}    
 \label{tbl:terminalA}
\end{table}

%%%%%%%%%%%%%%%%%%%%%%%%%%%%%%%%%%%%%%%%%%%%%%%%%%%%%%%%%%%%%%%%%%%%%%%%%%%%%%%%%%%%%
%
%     please remove the " % " symbol from \centerline{\includegraphics{fig01.eps}}
%     as it may ignore the figures.
%
%%%%%%%%%%%%%%%%%%%%%%%%%%%%%%%%%%%%%%%%%%%%%%%%%%%%%%%%%%%%%%%%%%%%%%%%%%%%%%%%%%%%%%

\section*{Data and Software Availability}

The Top-ML code can be found in {\href{https://github.com/XueGong-git/TopPep-ML}{https://github.com/XueGong-git/TopPep-ML}}. The datasets for antiCP 2.0 can be found in \href{https://webs.iiitd.edu.in/raghava/anticp2/download.php}{https://webs.iiitd.edu.in/raghava/anticp2/download.php}. The mACPpred 2.0 datasets are available from \href{https://github.com/nhattruongpham/mACPpred2}{https://github.com/nhattruongpham/mACPpred2}.

\section*{Author Contributions}

K.X. designed the experiments. J. T., J.W. and X.G. conducted the experiments and analyzed the results. J.T. drafted the initial manuscript. J.W. and X.G. revised the manuscript. All authors reviewed and contributed to the final manuscript.

\section*{Conflict of Interest}
The authors declare no competing financial interests.

%\appendix
\begin{acknowledgement}
J.T., J.W., and K.X. acknowledge the funding support from the Singapore Ministry of Education Academic Research Fund (MOE-T2EP20220-0010 and MOE-T2EP20221-0003). X.G. was supported by Nanyang Technological University under the Presidential Postdoctoral Fellowship grant 023545-00001.
\end{acknowledgement}

%%%%%%%%%%%%%%%%%%%%%%%%%%%%%%%%%%%%%%%%%%%%%%%%%%%%%%%%%%%%%%%%%%%%%

%%%%%%%%%%%%%%%%%%%%%%%%%%%%%%%%%%%%%%%%%%%%%%%%%%%%%%%%%%%%%%%%%%%%%

\bibliography{main}

\newpage % Starts a new page
\section*{TOC Image}

\begin{figure}[h]
    \centering
    \includegraphics[width=3.25in, height=1.5in]{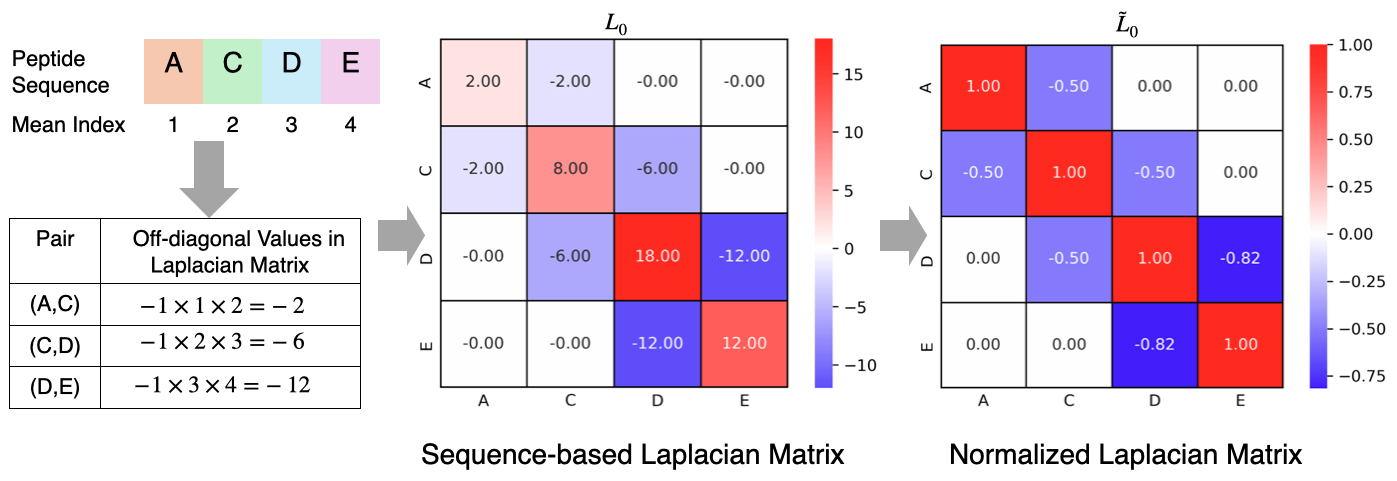} % Replace with your actual file name
\end{figure}

\end{document}